# Quantum Theory of a Steady State Laser

Frederick W. Cummings[*]

Michael T. Tavis[#]

## Abstract

A theory of Steady-State laser action is presented which loosely parallels the treatment by H. Fröhlich on energy storage in biological systems. The principal lasing elements, taken as N bio levels "molecules" and a single mode quantized radiation field are treated as a single quantum system. Laser action then is seen as a Bose-Einstein like condensation into the lowest energy mode of this correlated system when the pump power exceeds a certain critical value.

[*] Professor emeritus, University of California Riverside; present address: 611 Larkspur Plaza Dr., Larkspur, Ca.
[#] Retired-Casual,

## (I)  Introduction

The purpose of this paper is to discuss certain aspects of the theory of an idealized steady state laser. The principal elements involved, n two-level "molecules" (TLMs) interacting with a single-mode quantized radiation field, are treated as a single quantum system, following the solutions previously given. [1,2,3] Dicke[4] originally stressed the importance of treating such a system as a single correlated system; we will make use of the solutions to this problem as the "core" of the laser, and then bring this quantum system into interaction with a heat bath, "pump" and cavity loss mechanism by closely following the analysis of Fröhlich[5] on energy storage in Biological systems. In his paper, Fröhlich pointed out the interesting possibility that several properties of a living cell (e.g. their extraordinary dielectric properties, and the mechanism of cell division), may be understood by showing that when a certain critical rate of supply of energy to the cell is exceeded, the energy will not be completely thermalized but will be selectively stored in the lowest energy state of a collective longitudinal electric mode by a process strikingly akin to the familiar Bose-Einstein condensation. In the present paper, then, we will be viewing the onset of laser action as the analogue of Fröhlich's mechanism, an onset of laser action will be seen as a Bose-Einstein condensation phenomena into the lowest energy state of the collective system (TLM plus radiation field) when the pump power exceeds a certain critical value. This lowest state has a Gaussian distribution in photon number centered at the average photon occupation number. We begin by briefly reviewing certain aspects of the solution to the idealized TLM plus field problem.

II) <u>N Two-Level Molecules (TLMs) Interacting with a Single Mode Quantized Radiation Field.</u>

We now review briefly the relevant aspects of an idealize problem which will form the 'backbone" of our laser model. Let N non overlapping two-level systems which are all at the same resonant frequency $\omega$ interact with a single mode quantized radiation field, also of frequency $\omega$. The 'counter-rotating' interaction terms will be neglected and the TLMs are all taken to be at equivalent mode positions. The Hamiltonian (in units of $\hbar\omega$) is taken to be

$$H = R_3 + a^\dagger a - \kappa a R_+ - \kappa^* a^\dagger R_- \qquad \text{II-1)}$$

where $\kappa$ is the (complex) interaction constant, and

$$[a, a^\dagger] = 1 \qquad [R_+, R_-] = R_3 \qquad \text{and} \quad \kappa = |\kappa| e^{i\phi_1} \qquad \text{II-2)}$$

The zero of energy for each TLM is taken as midway between its excited energy, $\hbar\omega/2$, and its energy when unexcited, $-\hbar\omega/2$. States of the non-interacting system ($\kappa \to 0$) are defined such that

$$R_3 |r,m> = m|r,m> \qquad \text{II-3a)}$$

$$R_\pm |r,m> = e^{\pm\phi_2}[r(r+1) - m(m \pm 1)]^{1/2} \qquad \text{II-3b)}$$

$$R^2 |r,m> = \left[R_3^2 + \frac{R_+ R_- + R_- R_+}{2}\right]|r,m> = r(r+1)|r,m> \qquad \text{II-3c)}$$

$$a|n> = \sqrt{n} e^{i\phi_3}|n-1> \qquad \text{II-3d)}$$

The "cooperation number" r is a positive integer or half integer, and is analogous to the total angular momentum quantum number of spin systems, and m, the energy eigenvalue of non-interacting TLMs, has the range $-r \leq m \leq r \leq N/2$, where N is the total number of TLMs. Since the cooperation operator $R^2$ as well as the non-interacting energy operator $R_3 + a^\dagger a$ both commute with H, we can label eigenstates of H by the eigenvalues of $r(r+1)$ and c, eigenvalues of $R^2$ and $R_3 + a^\dagger a$ respectively. That is $c = <R_3> + <a^\dagger a>$, is a conserved quantity in the interacting system. For a given r and c, there are $2r+1$ degenerate states of the non-interacting system which are split up by the interaction into $2r+1$ discrete levels.

An eigenstate of H then has the expansion in the $|n>|r,m>$ basis

$$|r,c,j> = \sum_{n=max[0,c-r]}^{c+r} A_n^{(r,c,j)} |n>|r,c-n> \qquad \text{II-4a)}$$

and

$$H|r,c,j> = \lambda_{rcj}|r,c,j> \qquad \text{II-4b)}$$

Where j takes on the $2r+1$ values $-r \leq j \leq r$. Note that in (1), we assigned j values between 0 and 2r or between 0 and r+c depending on c greater than r or less than r and we defined effective eigenvalues of q. The exact solution to this problem is discussed in 1) and 3), and three distinct accurate approximation schemes for the eigenstates and eigenvalues are given in 2) for regions $c \ll 0$, $c \gtrsim r$ (where all states and eigenvalues are accurately given) and for all c, $r \gg 1$ (where only the ground and first few excited states are given accurately). In every case the ground state (j = -r) for r,c$\gg$ 1 is very closely a Gaussian distribution centered at $<n> = n_o$ for the probability of finding n photons in that state, namely

$$|A_n^0|^2 = C exp\left\{-\frac{(n-n_o)^2}{2\sigma^2}\right\}, \quad \text{II-5)}$$

where C is a constant and

$$\sigma^2(n) = \langle(n-n_o)^2\rangle = \frac{\sum|A_n^0|^2(n-n_o)^2}{\sum|A_n^0|^2}. \quad \text{II-6)}$$

In the references cited, it is found that[1]

$$\sigma^2 \cong \frac{q_o}{2\sqrt{\alpha_2}}, \quad \text{II-7)}$$

with $q_o$ being the so called effective eigenvalue of the ground state corresponding to j=-r, i.e.

$$q_o = \frac{c - \lambda_{j=-r}}{|\kappa|}. \quad \text{II-8)}$$

Without too many approximations it can be shown that

$$\sigma^2(n) = \langle n^2 \rangle - \langle n \rangle^2 = \frac{1}{2}\left[\frac{n_o[r^2 - (n_o - c)^2]}{3n_o - 2c}\right]^{1/2} \quad \text{when c>r>1,} \quad \text{II-9)}$$

$$n_o = \frac{2}{3}\left(c + \frac{1}{2}\right) + \frac{1}{3}\sqrt{3r^2 + 3r + \frac{3}{4} + c^2 + c + \frac{1}{4}} \cong \frac{2}{3}c + \frac{1}{3}\sqrt{3r^2 + c^2}, r, c \gg 1 \quad \text{II-10)}$$

Besides the ground state, in the region $c \gtrsim r \gg 1$, it is found that[2] the eigenvalues are linear in j,

$$\lambda_j = c + 2j|\kappa|\sqrt{n_o}, \quad -r \leq j \leq r \quad \text{II-11)}$$

The excited states also bear a strong resemblance to a discrete version of the familiar harmonic oscillator wave function, centered at <n> = $n_o$.

For Eq. II-9, the dispersion is of order $n_o$ except for the region r<<c. For example, when r>>c, $\sigma^2$~$n_o/\sqrt{6}$ and when r = c, $\sigma^2$= $n_o/\sqrt{12}$. The excited states for c >> 1 also have $<n_j>$ near c but the exact deviation of $<n_j>$ from c depends on j; the highly excited states have a positive average for the energy in the molecular system,

---

[1] This is reported incorrectly in reference (1). $\sigma^2 = \frac{n_o}{1 + q_o n_o/2\sqrt{\alpha_2}}$ there should be replaced by $\sigma^2 = \frac{q_o}{2\sqrt{\alpha_2}}$.

$$< n_j > = c - < m_j > \text{ where } 0 < < m_j > \ll < n_j > \text{ for } c \gg 1$$

III) Some Expressions Valid in Thermal Equilibrium

Dicke[4] considered a system of N2-level molecules in thermal equilibrium with one another. The states of the system |r,m> have an energy given by mE where E is the energy level separation of one two-level molecule. The states have a degeneracy given by

$$P(r) = \frac{N!(2r+1)}{(N/2+r+1)!(N/2-r)!}, \qquad \text{III-1)}$$

which is the same degeneracy found for the exact solutions for states $|r, c, j >$ having cooperation number r[3].

The average value of molecular energy in thermal equilibrium is ($\beta = E/kT$)

$$\overline{m} = m_{AV} = \frac{-N}{2} Tanh\left(\frac{\beta}{2}\right) \cong \frac{-NE}{4kT} \text{ for } \beta < 1, \qquad \text{III-2)}$$

and the standard deviation is

$$\sigma^2(m) = (m^2)_{AV} - m_{AV}^2 = \frac{N}{4} - \frac{m_{AV}^2}{N}. \qquad \text{III-3)}$$

The average value of r(r + 1) for a given value of the molecular energy (m, say) is[3]

$$\overline{[r(r+1)]} = \sum_m^{N/2} P(r)r(r+1) = \frac{N}{2} + m^2 \qquad \text{III-4)}$$

We also note that, for a given value of m, the mean square deviation in r(r+1) in terms of m is

$$\sigma(r(r+1)) = \overline{(r(r+1))^2} - \overline{r(r+1)}^2 = \frac{N^2}{4} - m^2. \qquad \text{III-5)}$$

Note that III-4) and III-5) are not thermal averages but are averages over the degeneracy. One could substitute the value of m̄ = $m_{AV}$ into III-4) and III-5) obtaining results associated with the thermal mean of m; however, the true thermal mean of $[r(r+1)]$ is obtained from III-4) as

$$< \overline{[r(r+1)]} >_{AV} = \frac{N}{2} + m^2{}_{AV} = \frac{3N}{4} + m_{AV}^2\left(1 - \frac{1}{N}\right), \qquad \text{III-6)}$$

which can be larger than express in III-4) when N large unless, as discussed below, $m_{AV}^2 \gg N$, in which case the two expressions are essentially identical when m is set equal to $m_{AV}^2$ in III-4). The mean square thermal deviation of r(r+1) is also given by

$$< \overline{[r(r+1)]^2} >_{AV} - < \overline{[r(r+1)]} >_{AV}^2 \qquad \text{III-7)}$$

$$= \frac{N(N-1)}{8} + \frac{(N-1)(N-2)}{N} m_{AV}^2 - \frac{2(2N-3)(N-1)}{N^3} m_{AV}^4$$

As an aside, the thermal average of III-5) is

$$< \sigma(r(r+1)) >_{AV} = \langle \frac{N^2}{4} - m^2 \rangle_{AV} = \frac{N(N-1)}{4}\left(1 - \frac{4m_{AV}^2}{N^2}\right) \qquad \text{III-8)}$$

so that if $m^2_{AV} \gg N \gg 1$, then the percent deviation from the mean of m is small decreasing as $1/\sqrt{N}$. The percent deviation from the mean of r(r +1) is also small but not decreasing as $1/\sqrt{N}$ but instead approximately as the square root of $m^2_{AV}/N$. Thus the mean of r(r+1) is approximately the smallest value compatible with the mean value of m, i.e. $r_{AV} \sim m_{AV}$. We expect this last property to be valid even when thermal equilibrium does not pertain.

IV) Introduction of Heat Bath, Pump and Cavity Loss

As mentioned in the introduction, idealized TLM + field systems of II) will be brought into interaction with a heat bath at temperature T with which can interchange energy. In the absence of external energy supply, or pump, and cavity losses, the heat bath will have the effect of bringing the system TLM + field into thermal equilibrium with it. In a helium-neon gas laser for example, the heat bath represents thermal collisions of the lasing neon atoms with helium atoms, other neon atoms, the walls of the enclosure and interaction with thermal radiation. We will be interested in steady state operation only, and the operating temperatures will be quite high, of the order of several thousand degrees Kelvin, although this fact is not critical to the present discussion. The presence of the pump and cavity loss will destroy the thermal equilibrium situation. In fact what we will show, following Fröhlich's[5] analysis, is that when the pump rate exceeds the cavity loss rate, by a certain critical amount, the system will undergo a kind of Bose-Einstein condensation into the lowest state j = -r of the quantum system of II). We represent the total combined system schematically as shown in figure 1.

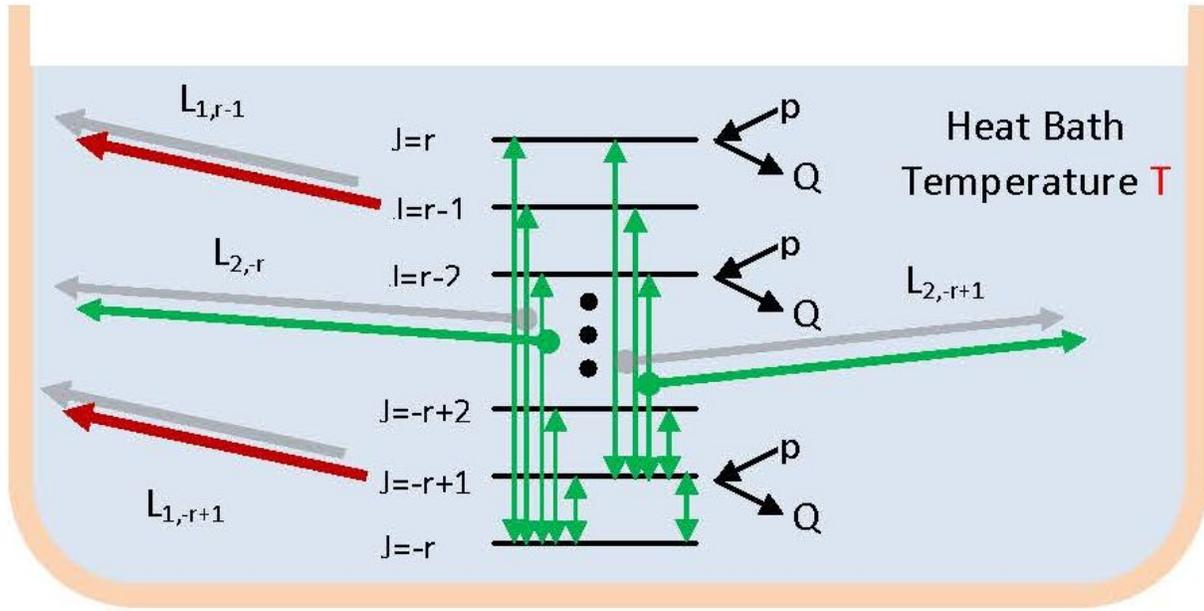

**Figure 1: Combined System including Eigenstates of TLM combined with Quantized Radiation Field with individual levels pumped by p , with external loss Q and in contact with a Thermal Bath at Temperature T with losses to the bath in first order $L_{1,j}$ and second order losses of $L_{2,j}$**

Here p represents the average rate at which energy is being supplied (equally) to each TLM. This is equivalent to energy being supplied equally to each level of the correlated system. Q will represent the cavity loss rate and will be taken as being the same loss rate from each level.

Prior to turning on the interaction ($\kappa =0$) the 2r + 1 levels of figure 1 will all be degenerate with energy cℏω, and the average of c will be given simply by the thermal distribution of <n> photons plus that of N two level atoms. It is assumed that the resonant frequency for the TLMs is the same as the mode frequency as stated above. As we turn on the interaction ($\kappa \neq 0$), but without a pump, this thermal energy will distribute itself over the 2r + 1 levels each with the same average c. We may write then

$$c = \frac{1}{2r+1} \sum_{j=-r}^{+r} <n_j> + <m_j> \qquad \text{IV-1)}$$

Now, for a given level, say "l", the $<m_l>$ will be determined by the quantum solutions discussed in II) once the $<n_l>$ is specified. So that we may write IV-1) as

$$(2r+1)[c-<m>] = \sum_{j=-r}^{+r} <n_j> \equiv \eta \qquad \text{IV-2a)}$$

$$<m> = \frac{1}{2r+1} \sum_{j=-r}^{+r} <m_j>. \qquad \text{IV-2b)}$$

The brackets are used here to represent thermal averages and are not to be confused with a strictly quantum average; double brackets would be more precise but clutter the notation. Now without the pump, we may presume that because of the usual arguments, $<n_j>$ will be given by the Planck formula,

$$< n_j >_T = \frac{1}{e^{\omega_j \beta} - 1} \text{ where } \beta = \frac{1}{kT} \qquad \text{IV-3a)}$$

and $\omega_j$ ($\hbar = 1$) represents the energies of the $2r + 1$ levels of figure 1. If $\kappa \to 0$, then $\omega_j \to \omega$ and

$$c \to \frac{1}{e^{\omega \beta} - 1} + <m> = \frac{1}{e^{\omega \beta} - 1} - \frac{N}{2} \tanh\left(\frac{\omega \beta}{2}\right). \qquad \text{IV-3b)}$$

Also, in the limit in which $c \gg r \gg 1$, we find from II-11) that $\omega_j = \omega + j\omega|\kappa|/\sqrt{c}$ and $\omega_j \sim \omega$ (In this sentence the c stands for the good quantum number c of section II and not the c of equation IV-1). In this case the average energy resides predominately in the radiation mode and $c \cong \frac{1}{e^{\omega \beta} - 1}$ as expected.

When the pump is turned on and p-Q>0, the $<n_l>$ will no longer be given by the Planck formula. Assume that the heat bath exchanges energy in quanta $\hbar\omega$ with the levels j at a rate which is nearly independent of $\omega$. The net rate of loss $L_{1l}$ of the level with energy $\omega_l$ and containing $<n_l>$ quanta can then be written in the form[5]

$$L_{1l} = \phi(T)(< n_l > e^{\omega_l \beta} - (1 + < n_l >)), \qquad \text{IV-4)}$$

where $\phi(T)$ may depend on temperature. In a higher order, individual units could exchange two or more quanta with the heat bath but never a fraction of a quantum. In second order, absorption of a quantum $\omega_l$ in conjunction with emission of a quantum of energy $\omega_j$, or vice versa, permits an exchange of energy between the levels and the heat bath in a range

$$0 < |\omega_l - \omega_j| \leq (\omega_r - \omega_{-r})$$

The net rate of loss $L_{2l}$ of the level l due to such processes can be written in the form

$$L_{2l} = \chi(T) \sum_{j=-r}^{+r} \left[< n_l > (1+ < n_j >) e^{(\omega_l - \omega_j)\beta} - < n_j > (1 + < n_l >)\right]. \qquad \text{IV-5)}$$

The general forms IV-4) and IV-5) are dictated by the requirement that in the absence of net external energy supply, or loss, p-Q = 0, thermal equilibrium ($L_{1l} = 0$, $L_{2l} = 0$) demands a Planck distribution for $<n_l>$. A slightly more general form for $L_{2l}$ would allow dependence of $\chi$ on $\omega_l$ or $\omega_j$ besides its temperature dependence.

The condition for a stationary state requires for each $l$

$$p = L_{1l} + L_{2l} + Q, \qquad \text{IV-6)}$$

where $p$ is the pump rate, assumed the same for each level l, and $Q$ is the qavity loss rate for each level. In the helium-neon laser, p is due to the exchange of energy from the metastable excited helium atoms to the lasing Neon atoms, and Q represents loss of energy from the resonant cavity via spontaneous emission into other free-space modes or through the finite reflectivity of the end mirrors, the latter being the dominant mechanism when the device is above threshold.

The average number of photons in state l is given by[5]

$$<n_l> = \left(1 + \frac{s}{\phi + \chi\eta}\right)\frac{1}{Ae^{\omega_l\beta} - 1} \quad \text{where } s = p - Q, \qquad \text{IV-7)}$$

and $\eta$ is defined in IV-2). Note that in Frohlich's paper N was used in place of $\eta$ and was not necessarily the thermal average. Also

$$A = \frac{\phi + \chi\sum_j(1+<n_j>)e^{-\omega_j\beta}}{\phi + \eta\chi} > 0. \qquad \text{IV-8)}$$

Introducing (following Fröhlich)

$$S(T) = s\sum_j e^{-\omega_j\beta}, \qquad \text{IV-9)}$$

One finds

$$S(T) = \phi\sum_j[<n_j> - (1+<n_j>)e^{-\omega_j\beta}], \qquad \text{IV-10)}$$

is independent of $\chi$, and A can then be written as

$$A = 1 - \frac{\chi}{\phi + \eta\phi}\left(\frac{S(T)}{\phi}\right) \leq 1. \qquad \text{IV-11)}$$

Now S=0 leads us back to A = 1 and thermal equilibrium in III-7). A $\leq$ 1 implies that, together with the condition $<n_l> \geq 0$,

$$A = e^{-\mu\beta} \text{ where } \omega_{-r} > \mu \geq 0. \qquad \text{IV-12)}$$

The point is now that a Bose-Einstein type of condensation will take place into the lowest state j = -r when μ approaches $\omega_{-r}$ very closely. As the parameter s increases (as the pump rate increases), S(T) will increase proportionally to s, so that A will decrease and μ will increase, eventually approaching $\omega_{-r}$ closely from below according to IV-12)

To see this more clearly we write, from IV-2) and IV-7)

$$\eta = \sum_{l=-r}^{+r}<n_l> = \left(1 + \frac{s}{\phi + \eta\chi}\right)\frac{1}{e^{(\omega_{-r}-\mu)\beta} - 1} + \left(1 + \frac{s}{\phi + \eta\chi}\right)\sum_{l\neq -r}^{+r}\frac{1}{e^{(\omega_l-\mu)\beta} - 1} \equiv <n_c> + n_n. \qquad \text{IV-13)}$$

An inequality involving $n_n$ is

$$n_n \leq \left(1 + \frac{s}{\phi + n_n\chi}\right)\sum_{l\neq -r}^{+r}\frac{1}{e^{(\omega_l-\omega_{-r})\beta} - 1}, \qquad \text{IV-14a)}$$

or,

$$\frac{n_n(\phi + n_n\chi)}{(\phi + n_n\chi + s)} \leq \sum_{\substack{l \neq -r}}^{+r} \frac{1}{e^{(\omega_l - \omega_{-r})\beta} - 1}. \qquad \text{IV-14b)}$$

Now the total η, the total average number of photons in all levels, is quite generally determined by s and ϕ, and will exceed the maximum imposed on $n_n$ by IV-14b). Introducing the excess number $e_l$ of quanta in $\omega_l$ over the number in thermal equilibrium

$$e_l \equiv <n_l> - <n_l>_T, \quad \sum_l e_l = \eta - \eta_T$$

Where $\eta_T$ is the total number of quanta in thermal equilibrium, on finds from IV-10) and IV-3),

$$S(T) = \phi \sum_l e_l(1 - e^{-\omega_l\beta}) \simeq \phi(1 - e^{-\varpi\beta})(\eta - \eta_T), \qquad \text{IV-15)}$$

where

$$\omega_{-r} \leq \varpi \leq \omega_r. \qquad \text{IV-16)}$$

Similarly from III-9) one has

$$S(T) \simeq s(2r + 1)e^{-\varpi\beta}. \qquad \text{IV-17)}$$

Then

$$\eta = \eta_T + \frac{(2r+1)s}{\phi(e^{\varpi\beta} - 1)}. \qquad \text{IV-18)}$$

Now ϖ will vary slightly with s in view of the limitation imposed by IV-16). Thus according to IV-18), η will increase linearly with s, until a value $s_o$ is reached above which $\eta \gtreqqless n_n$, since $n_n$ obeys IV-14). The inequality IV-14a) shows that as s gets very large, the maximum $n_n$ will increase only as the square root of s, i.e.,

$$n_n^{max} \to \sqrt{\frac{Bs}{\chi}}, \qquad \text{IV-19a)}$$

where

$$B = \sum_{\substack{j \neq -r}}^{r} \frac{1}{e^{(\omega_j - \omega_{-r})\beta} - 1}. \qquad \text{IV-19b)}$$

Thus since η increases directly proportional to s, laser action will set in as a condensation phenomena into the lowest state given by II-4) and II-9). Above some value $s_o$, the energy will selectively go into the state j = -r and $n_c$ (IV-13) will increase accordingly, since μ will approach $\omega_{-r}$ closely. The energy stored in the molecules is very closely zero for this highly populated ground state (as well as for the low lying excited states) but the average molecular energy stored in all of the states j, $<m> = \sum_j <m_j>$ will be a positive quantity.

Comparing IV-18) and IV-13) then gives for the average number of photons in the lasing mode,

$$<n_c> = \eta_T - \eta_n + \frac{(2r+1)s}{\phi(e^{\varpi\beta}-1)}. \qquad \text{IV-20)}$$

Far above threshold s>>1, we may find the dispersion in photon number in the laser mode from II-9) by the identification

$$\eta_c \to \eta_o = s\eta_T/\phi(T), \qquad \text{IV-21)}$$

where we have neglected $\eta_T - \eta_n$ in IV-20).

The threshold rate $s_o$ may be found approximately by supposing that $\eta \approx \eta_n$ and taking

$$\eta_T \simeq \frac{(2r+1)}{(e^{\varpi\beta}-1)}, \qquad \text{IV-22)}$$

and letting $\eta_n$ be approximated by taking the equality in IV-14a). Then we find that, after some straightforward algebra,

$$s_o = \frac{\phi}{\eta_T^2}\left[\left(\eta_T + \frac{2\phi}{\chi}\right)(2B - \eta_T)\right]. \qquad \text{IV-23)}$$

IV <u>Discussion</u>

It is pleasant to see laser action occurring as an example of an already familiar condensation mechanism, and to see it as being in direct correspondence to a promising model of biological energy storage; the feeling is strong that lasers should most properly be treated by considering the lasing atoms and radiation field as a single quantum system and we see in the above that a relatively simple view emerges.